\documentclass[journal=ancac3]{achemso}
\setkeys{acs}{articletitle = true}
\setkeys{acs}{etalmode = truncate}
\setkeys{acs}{maxauthors = 50}
\usepackage[version=3]{mhchem} 
\usepackage{amssymb,ulem}
\usepackage{amsmath,xcolor,bbm}
\usepackage{multirow}
\usepackage[left]{lineno}

\DeclareFontFamily{OMS}{oasy}{\skewchar\font48 }
\DeclareFontShape{OMS}{oasy}{m}{n}{%
         <-5.5> oasy5     <5.5-6.5> oasy6
      <6.5-7.5> oasy7     <7.5-8.5> oasy8
      <8.5-9.5> oasy9     <9.5->  oasy10
      }{}
\DeclareFontShape{OMS}{oasy}{b}{n}{%
       <-6> oabsy5
      <6-8> oabsy7
      <8->  oabsy10
      }{}
\DeclareSymbolFont{oasy}{OMS}{oasy}{m}{n}
\SetSymbolFont{oasy}{bold}{OMS}{oasy}{b}{n}
\DeclareMathSymbol{\smallleftarrow}     {\mathrel}{oasy}{"20}

\author{Siamak Khorasani}
\affiliation
{Department of Materials Science and Engineering, University of Washington, Seattle, WA}
\altaffiliation{Contributed equally to this work}
\author{Marc R. Bourgeois}
\affiliation
{Department of Chemistry, University of Washington, Seattle, WA}
\altaffiliation{Contributed equally to this work}
\author{David J. Masiello}
\email{masiello@uw.edu}
\affiliation
{Department of Materials Science and Engineering, University of Washington, Seattle, WA}
\alsoaffiliation
{Department of Chemistry, University of Washington, Seattle, WA}
\title[]
{Conservation of Pseudoangular Momentum in the Radiative Emission and Optical Excitation of Valley-Polarized Surface Lattice Resonances}

\begin{document}
 \begin{tocentry}
 \centering
 \includegraphics{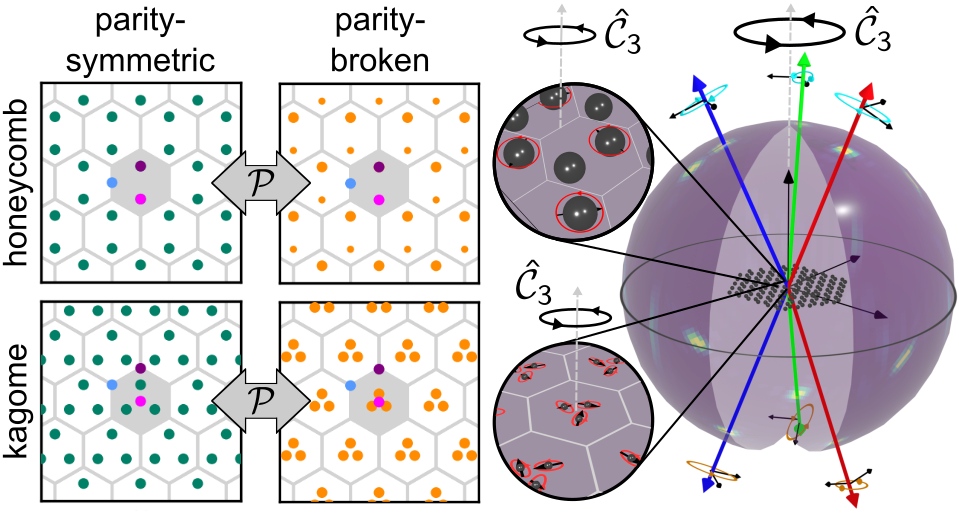}
 \end{tocentry}

\begin{abstract}
Surface lattice resonances (SLRs) are collective polaritonic excitations in nanoparticle arrays, with band-edge states enabling symmetry-based control of radiation including scattering, photoluminescence, conventional and polariton lasing, and condensation. Here, we show that the integer pseudoangular momentum (PAM) of valley-polarized SLRs in parity-broken hexagonal arrays of honeycomb and kagome arrangements is directly encoded in the phase and polarization structure of their emitted electromagnetic fields. \textcolor{black}{The effects of parity-symmetry breaking upon SLR PAM and the localized surface plasmon angular momentum associated with the individual nanoparticles composing the array are further investigated as are their radiative far-field signatures.} Leveraging electromagnetic reciprocity, we further propose a structured optical field possessing the same array symmetries and integer PAM as the underlying SLRs and demonstrate theoretically and numerically that it selectively couples to valley-polarized SLRs with matching quantized PAM. \textcolor{black}{Our results establish PAM-resolved far-field excitation and detection of polaritonic lattice resonances as a new platform for symmetry-selective nanophotonic control, chiral light generation, and angular-momentum-engineered metasurfaces, as well as for other applications leveraging valley pseudospin degrees of freedom in quantum information science.}
\end{abstract}

Valley degrees of freedom in hexagonal solid-state crystals enable symmetry-based approaches for controlling excitations associated with inequivalent $\text{K}$/$\text{K}'$ points, giving rise to valley-selective excitations \cite{valleyexcitongong2018nanoscale,valleyexcitonberghauser2018inverted, manca2017enabling, gendioneliu2023controlling}, valley-contrasting transport \cite{valleytransportwu2019intrinsic,valleytransportxiong2026observation, valleytransporthe2019silicon, valleytransportwojciechowska2025twist}, chiral atomic \cite{chiralatomiczhu2018observation, chiralatomizhu2024creating, chiralatomijuraschek2025chiral, chiralatomkim2016chiral, chiralatomlodahl2017chiral} and electronic motion, and suppressed intervalley scattering \cite{intervalleyscatteringvalleyzhang2025crystal, intervalleyscatteringvalleydevicemuis2025broadband}, thereby providing great potential for valleytronics \cite{valleytronicsmak2018light, valleytronicsschaibley2016valleytronics} and valley phononics  \cite{valleyphononicstian2020dispersion}. Valley excitations possessing three-fold rotational symmetry in hexagonal crystals are labeled by a pseudoangular momentum (PAM) quantum number in addition to their Bloch momentum and energy \cite{zhang2015chiral}, forming a set of quantities that are conserved between matter and radiation. While electronic and phononic PAM degrees of freedom have been extensively investigated in bulk \cite{chen2021propagating, ishito2023truly, ueda2023chiral,  minakova2026observation}, layered van der Waals \cite{chen2015helicity, unuchek2019valley, lian2023exciton, mi2025chiral}, and monolayer crystalline materials \cite{zhang2015chiral, chiralatomiczhu2018observation}, direct experimental access remains challenging \cite{zhu2018observation,bourgeois2025strategy,chiralatomijuraschek2025chiral}.


Two-dimensional periodic arrays of plasmonic nanoparticles represent a synthetic class of crystalline material with engineered lattice symmetries supporting collective polaritonic excitations known as surface lattice resonances (SLRs) \cite{cherqui2019plasmonic, kravets2018plasmonic, Manjavacas2018, Zou2004_1} that have enabled analogous exploration of phenomena such as chiral phonons \cite{chiralatomijuraschek2025chiral, zhang2015chiral, chen2019chiral}, valley edge states \cite{saito2021valley}, and chiral light emission \cite{chiralemissionzhang2022chiral, chiralemissionchen2025observation, chiralemissioncarlon2019optically, lasingKoenderink2019metasurfaces, wang2017band, zhou2013lasing, yang2015real, Torma_Kpoint_2019}, as well as enhanced polariton transport \cite{yadav2020strongly,  jin2023enhanced} and condensation \cite{hakala2018bose, moilanen2021spatial, daggett2026many}. SLRs arise from the hybridization of the localized surface plasmons (LSPs) of each nanoparticle with the optical modes supported by the periodic array structure, and may also possess quantized PAM in hexagonal arrays. However, unlike electronic and phononic excitations in atomic crystalline materials, SLRs strongly radiate, carrying information to the far-field. Special importance is attached to band-edge SLRs at high-symmetry Bloch momenta (such as valleys) because they exhibit local maxima in the photonic density of states, which facilitate lasing and condensation processes \cite{zhou2013lasing, hakala2018bose, moilanen2021spatial, daggett2026many}. Despite prior investigations of polarized lasing emission from valley SLRs \cite{Torma_Kpoint_2019, lasingKoenderink2024spontaneous,lasingKoenderink2025metasurface,zheng2026circularly}, the connection between the polarization states of radiated photons to the underlying SLR PAM has remained unclear, hindering access to this polaritonic valley degree of freedom.

In this work, we establish the conservation of PAM in the radiative emission and optical excitation of valley-polarized SLRs in noncentrosymmetric honeycomb and kagome plasmonic arrays. The impact of parity- ($\hat{\cal P}$) symmetry breaking through continuous deformation of lattice geometry on SLR PAM and LSP angular momentum (AM) are investigated for both prototypical hexagonal lattices. Specifically, by examining the polarization, nonlocal phase profile, and angular distribution of the far-field radiation emitted by SLRs, we reveal the connection between the discrete rotational symmetries of the field and the PAM and AM of the sourcing SLR mode and its constituent LSP components, respectively. Further, by exploiting electromagnetic reciprocity, we invert this mapping between SLR and radiation field to design a structured optical excitation field that possesses the same crystalline symmetry (i.e., discrete translation and rotation symmetries) and integer PAM as the valley-polarized SLRs in the noncentrosymmetric hexagonal arrays studied. In contrast to circularly polarized light, we demonstrate theoretically and numerically that our proposed free-space field selectively couples to valley SLRs with matching quantized PAM, providing a new route toward symmetry-selective excitation of collective polaritonic lattice modes. \textcolor{black}{Together, these results expand possibilities for the excitation and detection of valley-polarized lattice resonances in open polaritonic/nanophotonic cavities by establishing the connection between shared symmetries expressed in material and radiated field polarization textures associated with K/K' points, complementing experimental approaches to leverage momentum-space wavefront polarization and phase information in related systems \cite{gerchberg1972practical,machfuudzoh2026cathodol, lasingKoenderink2024spontaneous, coenen2014directional, lasingKoenderink2025metasurface, chen2017general}.}


\begin{figure}
    \centering
    \includegraphics[width=1.0\linewidth]{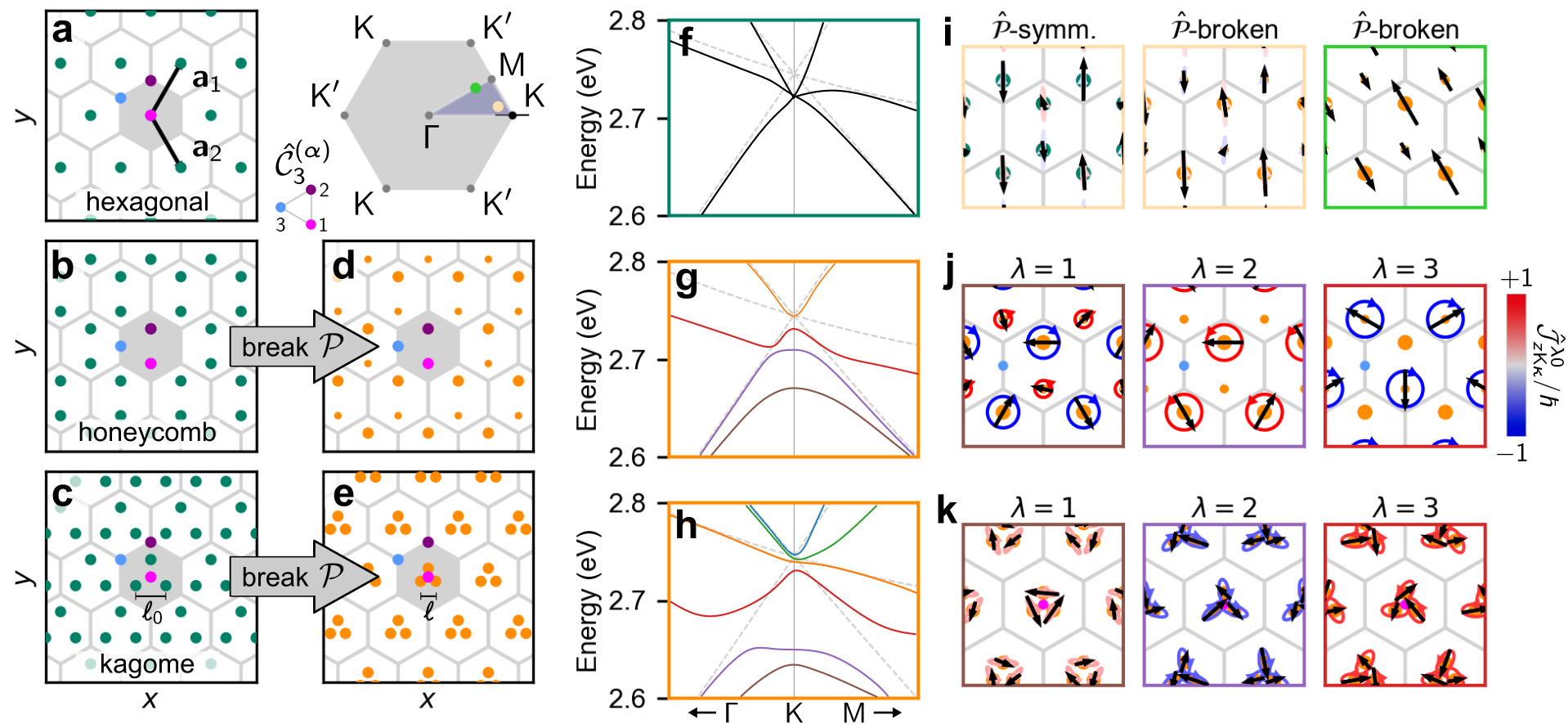}
    \caption{{\bf Parity breaking and angular momentum in 2D hexagonal arrays.} a) Direct space (left) and Brillouin zone (right)  representations of a 2D hexagonal lattice with basis vectors $\mathbf{a}_1$ and $\mathbf{a}_2$. b) and c) show parity-symmetric honeycomb and kagome array geometries. d) and e) show parity-broken honeycomb and kagome array geometries. Magenta, purple, and blue circles within the central unit cells in (a)-(e) indicate locations of $\hat{\mathcal{C}}^{(\alpha)}_{3}$ rotation axes. f)-h) Energy-momentum dispersion diagrams of SLRs in the hexagonal and parity-broken honeycomb and kagome arrays, respectively, in the vicinity of the $\text{K}$ point spanned by the black line in (a). Specific details regarding array and nanoparticle geometries for (f)-(h) can be found in the main text. i) LSP dipole polarizations at an arbitrary time (black arrows) for the $\lambda=1$ SLRs of the parity-symmetric and -broken honeycomb arrays at the Bloch momenta indicated by the light tan and green markers located inside and on, respectively, the boundary of the irreducible BZ shaded blue in (a). The LSP polarization ellipses traced by the tip of each arrow over an oscillation period are included and colored according to $\hat{\mathcal{J}}^{\lambda 0}_{z\text{K}\kappa}$. j) k) LSP dipole moment polarizations (black arrows) for $\text{K}$ point SLRs ($\lambda=1-3$) of the parity-broken honeycomb and kagome arrays from (g) and (h), respectively. LSP polarization ellipses are colored according to $\hat{\mathcal{J}}^{\lambda 0}_{z\text{K}\kappa}$.}
    \label{F1}
\end{figure}




The connection between geometric parity breaking and nonzero (pseudo-) angular momenta associated with Bloch wave SLRs in 2D hexagonal arrays is investigated in Fig. \ref{F1}  for lattices with $N_\kappa = 1-3$ nanoparticle sites per unit cell. Each 2D hexagonal array, independent of $N_\kappa$, is spanned by primitive lattice vectors $\mathbf{a}_1$ and $\mathbf{a}_2$ with $|\mathbf{a}_{1,2}| = a$.  Fig. \ref{F1}a shows the simplest 2D hexagonal array with $N_\kappa = 1$ in direct space (left). The Brillouin zone of the hexagonal array  in reciprocal space (independent of $N_\kappa$) is depicted to the right with $|\mathbf{K}_{\text{K}/\text{K}'}| = 4\pi/3a$. Figs. \ref{F1}b--c present the direct space geometries of the honeycomb and kagome arrays with  $N_\kappa =$ 2 and 3 identical sites, respectively. These centrosymmetric arrays are characterized by $6/mmm$ ($D_{6h}$) crystallographic symmetry. In addition to parity, the crystallographic point group $G_{\mathbf{0}}/T $, where the subscript indicates the $\Gamma$ point $\mathbf{K} = \mathbf{0}$, also contains three distinct three-fold rotation symmetry elements $\hat{\mathcal{C}}_3^{(\alpha)}$ ($\alpha=1,2, 3$), which are marked by color-coded circles in Figs. \ref{F1}a--c. When $N_\kappa > 1$ it is possible to modify the unit cell geometry to lower the crystallographic symmetry by breaking parity while maintaining the three-fold rotational symmetries. Fig. \ref{F1}d shows such a parity-broken honeycomb array, where $\hat{\mathcal{P}}$ breaking is achieved by placing inequivalent nanoparticles at sites $\kappa = 1,2$ as indicated by distinct orange site marker sizes. Fig. \ref{F1}e shows a parity-broken kagome lattice, where $\hat{\mathcal{P}}$ is broken by contracting the three equivalent sites around the unit cell center such that $l < l_0$, where $l_0= $ $a/2$ . Despite the different manifestations of symmetry breaking, the parity-broken (noncentrosymmetric) honeycomb and kagome arrays shown in Fig. \ref{F1} both have crystallographic symmetry $\bar{6}m2$ ($D_{3h}$). Although these noncentrosymmetric arrays do not possess parity symmetry, the out-of-plane reflection symmetry $\hat{\sigma}_h \in G_{\mathbf{0}}/T$ for all 2D monolayer arrays, precluding structural chirality in these systems.


Energy-momentum dispersions, lifetimes, and Bloch mode symmetries of collective SLRs arising from electromagnetically-coupled LSPs supported by individual nanoparticles positioned at lattice sites can be accurately determined using the coupled-dipole method (Supporting Information) \cite{cherqui2019plasmonic, kravets2018plasmonic, Markel1993, Zou2004_1, Manjavacas2018}. Figs. \textcolor{black}{\ref{F1}f--h} show the coupled-dipole energy-momentum dispersion diagrams of SLRs in parity-symmetric (green) and parity-broken (orange) hexagonal, honeycomb, and kagome arrays in the vicinity of the $\text{K}$ point spanned by the black line in Fig. \ref{F1}a. For all coupled-dipole calculations, the hexagonal lattice primitive vector magnitudes are $a = 780$ nm and the nanoparticle diameter $d = 100$ nm for all $\hat{\mathcal{P}}$-symmetric lattices, while the $\hat{\mathcal{P}}$-broken honeycomb lattice uses $d_1 = 110$ nm and $d_2 = 90$ nm. The intra-particle spacing is $l_0 = 390$ nm for the $\hat{\mathcal{P}}$-symmetric kagome lattice and $l= 195$ nm for the $\hat{\mathcal{P}}$-broken kagome lattice. The dispersion of SLRs supported by each array geometry qualitatively follows the dashed gray lines representing the empty-lattice dispersion \cite{guo2017geometry, cherqui2019plasmonic} characterizing the propagation of a free photon subject to Bloch wave boundary conditions in an otherwise empty unit cell.  The empty-lattice dispersion indicates where symmetry-related band crossings arise, though the inclusion of discrete nanoparticles within the unit cell may lift some or all of the degeneracy. Consequently, the existence of SLR band degeneracies is governed by the symmetry of the full nanoparticle lattice rather than by the empty-lattice construction alone, and must be established through a group-theoretical analysis of the relevant irreducible representations \cite{dresselhaus2007group}. In particular, at high-symmetry reciprocal space points along the perimeter of the irreducible BZ, $G_\mathbf{K}/T$ can contain additional symmetry elements beyond $\hat{\sigma}_h$ and the identity $\hat{E}$, which may introduce band degeneracies or connect SLR dispersions at equivalent Bloch momenta. At each of the $\text{K}$ ($\text{K}'$) points, for instance, $\hat{\mathcal{C}}_3 \in G_\mathbf{K}/T$ for both parity-symmetric and parity-broken array classes, enforcing identical band energies and related SLR polarizations at the three equivalent $\text{K}$ ($\text{K}'$) points. In the parity-symmetric arrays, $\hat{\mathcal{P}} \in G_\mathbf{0}/T$  and two-fold SLR band degeneracies -- along with the accompanying loss of well-defined polarization symmetries -- occur at the BZ corners. In the parity-broken arrays, however, the absence of $\hat{\mathcal{P}}$ from $G_\mathbf{0}/T$ lifts the remaining SLR band degeneracies at individual $\text{K}$ ($\text{K}'$) points as observed in Fig. \ref{F1}g,h.

In addition to lifting valley degeneracies, parity breaking generally introduces a net circulation of LSP polarization vectors associated with Bloch wave excitations at arbitrary $\mathbf{K}$ within the interior of the irreducible BZ, away from high-symmetry points and planes \cite{coh2023classification}. Within the coupled-dipole model, the electric dipole moment associated with the nanoparticle located at $\mathbf{x}_{\mathbf{n}\kappa} = \mathbf{x}_\mathbf{n} + \mathbf{r}_\kappa$, i.e., at site $\kappa$ within the $\mathbf{n}$th unit cell, comprising the SLR characterized by the two-dimensional Bloch vector $\mathbf{K}$ and in-plane band index $\lambda$, is denoted by $\mathbf{p}^{\lambda \mathbf{K}}_{\mathbf{n} \kappa}$. It satisfies the Bloch form $\mathbf{p}^{\lambda \mathbf{K}}_{\mathbf{n} \kappa} = \mathbf{p}^\lambda_{\mathbf{K} \kappa} \exp(i \mathbf{K}\cdot \mathbf{x}_\mathbf{n})$. In analogy to the AM associated with the atomic motion of phonons \cite{zhang2014angular, zhang2015chiral}, we express the $z$-component of the total AM associated with the precession of LSP dipoles in a given SLR as $J^{\lambda}_{z\mathbf{K}} = \sum_\kappa J^{\lambda}_{z\mathbf{K}\kappa}$, with ${J}^{\lambda}_{z \mathbf{K}\kappa} = (M_\kappa /q^{\lambda 2}_\mathbf{K})\sum_\mathbf{n} \{ \mathbf{p}^{\lambda \mathbf{K}}_{\mathbf{n}\kappa} \times \dot{\mathbf{p}}^{\lambda \mathbf{K}}_{\mathbf{n}\kappa}  \}_z$.
Quantities $q_\mathbf{K}^\lambda$ and $M_\kappa$ are the SLR effective charge and site  masses, respectively, as described in more detail in the Supporting Information. The finite lifetimes necessitated by the complex-valued SLR energies $\hbar \omega_\lambda$ require LSP AM and SLR PAM to decay with time. Nevertheless, suitably defined, the LSP AM at time zero, $J^{\lambda 0}_{z\mathbf{K}\kappa}$, behaves similarly to the AM arising in lossless settings with $\hbar \omega_\lambda \in \mathbb{R}$ \cite{zhang2014angular, zhang2015chiral}. \textcolor{black}{To connect the per-site AM to the geometries of the individual LSP orbits, we define a scaled $\hat{\mathcal{J}}^{\lambda 0}_{z\mathbf{K}\kappa} \propto J^{\lambda 0}_{z\mathbf{K}\kappa}$, ensuring that $\hat{\mathcal{J}}^{\lambda 0}_{z\mathbf{K}\kappa}/\hbar = \pm 1$ for all circular orbits independent of the number of sites $N_\kappa$ within the unit cell and the number of sites undergoing motion within a given SLR (see Supporting Information).}



The first two panels of Fig. \ref{F1}i depict snapshots of the LSP polarizations (black arrows) at an arbitrary time for the $\lambda=1$ SLRs of the parity-symmetric and parity-broken honeycomb arrays at the Bloch momentum indicated by the light tan marker in Fig. \ref{F1}a within the interior of the irreducible BZ. The LSP polarization ellipses traced by the tip of each arrow over an oscillation period are included and colored according to $\hat{\mathcal{J}}^{\lambda 0}_{z\mathbf{K}\kappa}$. As anticipated for a general $\mathbf{K}$ within the interior of the irreducible BZ \cite{coh2023classification}, the SLR polarizations at the light tan point in Fig. \ref{F1}i exhibit nonzero net LSP AM $\hat{\mathcal{J}}^{\lambda 0}_{z\mathbf{K}}$ ($J^{\lambda 0}_{z\mathbf{K}}$) for the parity-broken structure only. Although the individual site LSP polarizations are elliptical rather than linear irrespective of the presence/absence of parity, perfect cancellation of LSP polarizations occurs such that the net $\hat{\mathcal{J}}^{\lambda 0}_{z\mathbf{K}}=0$ for the parity-symmetric structure. This simple connection linking the presence of parity in the crystallographic point group to the existence of SLRs associated with nonzero net LSP AM does not generally apply along the perimeter of the irreducible BZ as discussed above \cite{coh2023classification, dresselhaus2007group}. This point is highlighted in the right-most panel of Fig. \ref{F1}i, which shows a snapshot of the LSP polarizations associated with an SLR with Bloch $\mathbf{K}$ marked by the green dot in Fig. \ref{F1}a along the $\Gamma\text{-M}$ direction. Despite the absence of parity, the inclusion of an additional mirror symmetry in $G_\mathbf{K}/T$ causes the individual LSPs to be linearly polarized and the net LSP AM to be zero. 

At the high-symmetry $\text{K}/\text{K}'$ valleys at the BZ corners of hexagonal arrays, $\hat{\mathcal{C}}_3  \in G_\mathbf{K}/T $ and the direct space SLR mode polarization vector fields $\{ \mathbf{p}^{\lambda \mathbf{K}}_{\mathbf{n}\kappa} \}$ are simultaneous eigenmodes of the $\hat{\mathcal{C}}_3$ rotation operator, i.e., 
\begin{equation}
    \begin{aligned}
        \hat{\mathcal{C}}_3 ^{(\alpha)} \bigg\{ \mathbf{p}^{\lambda \mathbf{K}}_{\mathbf{n}\kappa} \bigg\} &= e^{-i (2\pi/3) m^{(\alpha)}_{\lambda \mathbf{K}} } ~ \bigg\{ \mathbf{p}^{\lambda \mathbf{K}}_{\mathbf{n}'\kappa'} \bigg\}, \\
    \end{aligned} 
    \label{Cn_gen_eig}
\end{equation}
and the total PAM $m^{(\alpha)}_{\lambda \mathbf{K}} \in \{\pm1,0\}$ is integer valued. Eq. \eqref{Cn_gen_eig} states that $\hat{\mathcal{C}}_3$ rotation of the LSP displacement field at some arbitrary time $t_0$ results in a new snapshot of the same SLR displacement field, but at time $t = t_0 + \Delta t_\alpha$ with phase dictated by $m^{(\alpha)}_{\lambda \mathbf{K}}$. This additional symmetry requirement dictates LSP orbit geometries associated with SLRs as depicted in Fig. \ref{F1} panels (j) and (k) for parity-broken $N_\kappa=2$ and $N_\kappa=3$ arrays, respectively. With knowledge of the LSP orbit directions, the total PAM can be assigned upon inspection of dipole polarization snapshots before and after $\hat{\mathcal{C}}_3$ rotation according to $m^{(\alpha)}_{\lambda \mathbf{K}} = (3/T_\text{orb})\Delta t_\alpha$, where $T_\text{orb}$ is the orbit period. The total PAM of each $\text{K}$ valley SLR associated with the parity-broken arrays in Figs. \ref{F1}g and \ref{F1}h are tabulated in Table \ref{PAM_table}. The $\text{K}$ and $\text{K}'$ valley SLR PAM values are interrelated by $m^{(\alpha)}_{\lambda \text{K}'} = - m^{(\alpha)}_{\lambda \text{K}}$ \cite{zhang2015chiral}. Already, the distinction between LSP AM $\hat{\mathcal{J}}^{\lambda 0}_{z\mathbf{K}\kappa}$ and SLR PAM $m^{(\alpha)}_{\lambda \text{K}}$ is apparent in the comparison of PAM values from Table \ref{PAM_table} with LSP orbit directions (and eccentricities) associated with the valley SLRs presented in Figs. \ref{F1}j and \ref{F1}k. 

\begin{table}[ht] \centering \renewcommand{\arraystretch}{1.2} \begin{tabular}{|c|c||c|c|c|c|c|c|} \hline \multicolumn{8}{|c|}{K valley SLR PAM $m^{(\alpha)}_{\lambda \text{K}}$} \\ \hline \multicolumn{2}{|c||}{} & \multicolumn{6}{c|}{$\lambda=$} \\ \hline $N_{\kappa}$ & $\alpha$ & 1 & 2 & 3 & 4 & 5 & 6 \\ \hline 2 & 3 & 0 & $-1$ & $+1$ & 0 & -- & -- \\ \hline 3 & 1 & 0 & $+1$ & $-1$ & $-1$ & $+1$ & 0 \\ \hline \end{tabular}
\caption{PAM of K point SLRs for parity-broken arrays in Fig. \ref{F1}g,h with respect to $\hat{\mathcal{C}}_3^{(\alpha)}$ rotation axes, where $\alpha=3$ (blue) for $N_\kappa = 2$, and $\alpha=1$ (pink) for $N_\kappa = 3$. Bands of each array type are indexed by $\lambda$ according to increasing energy.}
\label{PAM_table} 
\end{table}

\begin{figure}
    \centering
    \includegraphics[width=1.0\linewidth]{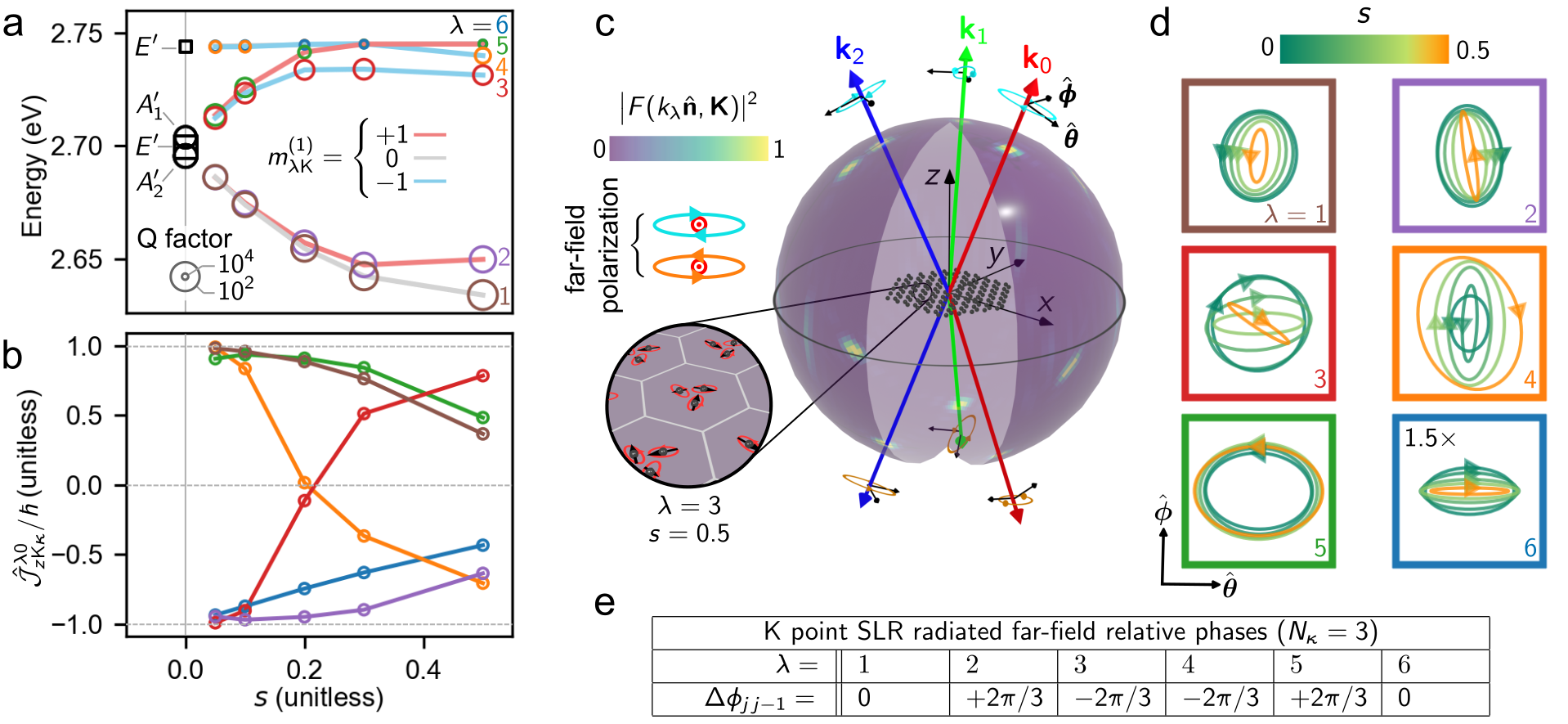}
    \caption{{\bf Evolution of LSP AM and radiated far-field polarization of $\text{K}$ point SLRs in kagome arrays under continuous symmetry breaking.} a) Real and imaginary parts of $\text{K}$ point SLR energies $\hbar \omega_\lambda$ of the $N_\kappa=3$ arrays as a function of symmetry breaking parameter $s$. Marker colors and sizes correspond to band indices $\lambda$ and SLR quality factors $Q$, respectively, while line colors denote PAM values $m^{(1)}_{\lambda \textrm{K}}$.  b) Scaled LSP AM $ \hat{\mathcal{J}}^{\lambda 0}_{z\text{K}\kappa} $ as a function of $s$. c) Angular distribution of radiated $|\mathbf{E}^{\lambda \mathbf{K}}_\text{rad}(\hat{\mathbf{n}}, \bar{\omega}_\lambda)|^2$ produced by the $\lambda=3$ $\text{K}$ valley SLR. \textcolor{black}{The color map on the spherical shell represents the normalized squared modulus of the structure factor $\big|F(k_{\lambda} \hat{\mathbf{n}}, \mathbf{K})|^2$ as a function of observation direction $\hat{\mathbf{n}}$ and $k_{\lambda}=\bar{\omega}_\lambda/c$. } d) Evolution of the far-field polarization state radiated into direction $\hat{\mathbf{k}}_0$  for each $\text{K}$ valley SLR as $s$ varies continuously within $0\leq s \leq 1/2$.  e) Relative phases $\Delta \phi_{j\,j-1}$ between radiation beamed into the three $\text{K}$ point directions $\hat{\mathbf{k}}_j$ \textcolor{black}{referenced with respect to the $\hat{\mathcal{C}}_3^{(1)}$ axis of the unit cell equidistant from each detector.}    }
    \label{F2}
\end{figure}

While Figure \ref{F1} focuses on the binary impacts of presence/absence of $\hat{\mathcal{P}}$ in $G_\mathbf{K}/T$ for arrays with hexagonal symmetry, Figure \ref{F2} investigates the relationship between near- and far-field observables as the degree of parity breaking is continuously varied. The degree of symmetry breaking in the $N_\kappa=3$ arrays can be quantified by defining the dimensionless parameter $s \equiv 1 - \ell/\ell_0$, where $\ell_0$ and $\ell$ are edge lengths of the equilateral triangles formed by the kagome sites in the parity-symmetric and parity-broken lattices, respectively. The parity-symmetric array is characterized by $s=0$, and a practical maximum $s_\text{max}\approx 0.75$ is imposed by the finite nanoparticle diameter. Fig. \ref{F2}a presents the evolution of $\text{K}$ point SLR energies, lifetimes, and PAM as $s$ varies between $0\leq s \leq 1/2$ for $N_\kappa = 3$ arrays with $a=780$ nm and $d = 100$ nm nanoparticle diameters. As $s$ increases from zero, each pair of doubly degenerate $E'$ SLRs (indicated by square markers) splits, and all six in-plane SLRs become labeled by the same one-dimensional irreducible representation. Dimensionless SLR quality factors $Q = \text{Re}\{\hbar \omega_\lambda\}/\text{Im}\{\hbar \omega_\lambda\}$ are used to quantify excitation lifetimes and are indicated by marker size in Fig. \ref{F2}a. As the site-to-site coupling within each unit cell increases with increasing $s$, two bands shift towards lower energy and lower $Q$ factors. Meanwhile, the energies of the remaining four bands increase and become compressed in the vicinity of the $K$ point Rayleigh anomaly, exhibiting more complicated $Q$ factor evolution.

The scaled LSP site AM $\hat{\mathcal{J}}^{\lambda 0}_{z\mathbf{K}\kappa}$ is presented in Fig. \ref{F2}b as a function of $s> 0$ for each of the $\text{K}$ valley SLRs supported by the parity broken kagome arrays. Two sets of triplets $\lambda=\{2, 3, 6\}$ (purple, red, blue) and $\lambda=\{1, 4, 5\}$ (brown, orange, green) are observed with qualitatively opposite site AM behavior with respect to $\hat{\mathcal{J}}^{\lambda 0}_{z\mathbf{K}\kappa} = 0$. Meanwhile, four modes ($\lambda=\{1, 2, 5, 6\}$) exhibit fixed handedness, defined as $\text{sgn}\big( \hat{\mathcal{J}}^{\lambda 0}_{z\mathbf{K}\kappa}\big)$, that decreases with increasing $s$ in the displayed range, while that of the other two ($\lambda=\{3, 4\}$) changes sign near $s \approx 0.2$. The impact of parity-breaking on the LSP AM in the case of honeycomb arrays is presented in the Supporting Information (SI Fig. S5). 
\textcolor{black}{
In general, the scaled site AM $\hat{\mathcal{J}}^{\lambda 0}_{z\mathbf{K}\kappa}$ determines the eccentricities of the LSP polarization ellipses characterizing each $\mathbf{p}^{\lambda \mathbf{K}}_{\mathbf{n}\kappa}$ (see Supporting Information). The PAM, meanwhile, fixes the relative polarization ellipse orientations and phases to satisfy Eq. \eqref{Cn_gen_eig}. Therefore, their distinct physical origins notwithstanding, the LSP AM and SLR PAM each contribute to the particular geometries of SLR polarization structures. 
}

The significant far-field electromagnetic coupling underlying SLRs in arrays simultaneously generates radiation beamed into specific directions based on the SLR dispersion \cite{cherqui2019plasmonic, kravets2018plasmonic}. For unit cell configurations possessing $\hat{\mathcal{C}}_n$ symmetry, the phase correlations between unit cells, as well as between LSPs within each unit cell, dictated by PAM quantization at certain $\mathbf{K}$ are imprinted on radiated electromagnetic fields. Notably, the band-edge states facilitating lasing and condensation processes tend to be located at the center and corners of BZs, e.g., at the valleys, where PAM becomes quantized. 
Within the coupled-dipole approximation \cite{Zou2004_1, Markel1993, cherqui2019plasmonic, Manjavacas2018}, the electric field radiated by the SLR mode with band index $\lambda$, in-plane wave vector $\mathbf{K}$, and energy $\textcolor{black}{\hbar \bar{\omega}_\lambda=\textrm{Re}\{\hbar\omega_\lambda\}}$ into direction $\hat{\mathbf{n}}$, with $\mathbf{x} = r\hat{\mathbf{n}}$, can be expressed as $\mathbf{E}^{\lambda \mathbf{K}}_\text{rad}(\mathbf{x}, \textcolor{black}{\bar{\omega}_\lambda}) = \sum_{\mathbf{n} \kappa} \bar{\mathbf{G}}_{0}(\mathbf{x}, \mathbf{x}_{\mathbf{n} \kappa}; \textcolor{black}{\bar{\omega}_\lambda})  \cdot \mathbf{p}^{\lambda \mathbf{K}}_{\mathbf{n} \kappa}$,
where $\bar{\mathbf{G}}_{0}(\mathbf{x}, \mathbf{x}_{\mathbf{n} \kappa}; \textcolor{black}{\bar{\omega}_\lambda})$ is the free space dyadic Green tensor. For observation points within the radiation zone, where $|\mathbf{x}| \gg |\mathbf{x}_{\mathbf{n} \kappa}|$, the Fraunhofer approximation can be invoked such that (Supporting Information) 
\begin{equation}
    \begin{aligned}
        \mathbf{E}^{\lambda \mathbf{K}}_\text{rad}(r\hat{\mathbf{n}}, \textcolor{black}{\bar{\omega}_\lambda}) &\sim F(k_{\lambda} \hat{\mathbf{n}}, \mathbf{K})\bar{\mathbf{G}}_0(\mathbf{x}, \mathbf{0};\textcolor{black}{\bar{\omega}_\lambda}) \cdot \mathbb{P}^\lambda_\mathbf{K}, \\
    \end{aligned}
    \label{angdist_FF}
\end{equation}
where $\textcolor{black}{k_{\lambda}=\bar{\omega}_\lambda/c}$ with $k_{\lambda}^2 = |\mathbf{K}|^2 + k^2_{\lambda z}$, $\bar{\mathbf{G}}_0(\mathbf{x}, {\bf 0};\textcolor{black}{\bar{\omega}_\lambda}) = \big[ k_\lambda ^2 \bar{ \mathbf{I}} + \nabla \nabla \big] \exp \big( {ik_\lambda r} \big)/r $, and the unit cell polarization is defined as $\mathbb{P}^\lambda_\mathbf{K} = \sum_\kappa \mathbf{p}^\lambda_{\mathbf{K}\kappa} \exp({-i k_\lambda \hat{\mathbf{n}} \cdot \mathbf{r}_\kappa})$. The structure factor $F(k_{\lambda} \hat{\mathbf{n}}, \mathbf{K}) = N^{-1/2} \sum_{\mathbf{n}}  \exp [ i ( \mathbf{K} - k_\lambda \hat{\mathbf{n}}) \cdot \mathbf{x}_{\mathbf{n}} ]$
captures the modulation of the angular distribution arising from diffraction, where $N$ is the number of unit cells employed under Born-von Karman boundary conditions. Fig. \ref{F2}c presents a scheme of a parity-broken kagome SLR radiating into the far-field region with the squared modulus of the structure factor $|F(k_\lambda \hat{\mathbf{n}}, \mathbf{K})|^2$ plotted on the spherical shell surrounding the array. The structure factor ensures each of the $\text{K}$ ($\text{K}'$) SLRs simultaneously beams radiations into the three equivalent $\text{K}/\text{K}'$ directions (indexed by $j\in\{0,1,2\}$) satisfying $k_\lambda \sin \theta  = K$. \textcolor{black}{Similar combined coupled dipole and group theoretical descriptions of valley SLRs and their associated radiation field characteristics have been shown to accurately predict experimentally measured far-field polarization characteristics of lasing emission from molecular emitters integrated into parity-symmetric honeycomb arrays \cite{guo2019lasing}.}



Equation \eqref{angdist_FF} explicitly relates the far-field polarization state of light radiated into direction $\hat{\mathbf{n}}$ to the orthographic projection of the planar 2D $\boldsymbol{\mathbb{P}}^\lambda_\mathbf{K}$ effective unit cell polarization state onto the $\hat{\boldsymbol{\varphi}}$-$\hat{\boldsymbol{\theta}}$ plane. The six panels in Fig. \ref{F2}d depict the evolution of the far-field polarization ellipses, in the $(E_\theta, E_\varphi)$ basis, characterizing the polarization state of light radiated into direction $\hat{\mathbf{k}}_0$ for each $\text{K}$ point SLR as $s$ varies continuously within $0\leq s \leq 1/2$. The inequivalence between the far-field polarization states and any of the individual LSP polarization states participating in the SLR is responsible for the qualitatively distinct evolutions of LSP and radiated photon polarizations evident in Figs. \ref{F2}b and \ref{F2}d. Incidentally, it is the orthographic projection of the planar 2D $\boldsymbol{\mathbb{P}}^\lambda_\mathbf{K}$ onto the $\hat{\boldsymbol{\varphi}}$-$\hat{\boldsymbol{\theta}}$ planes in the $\pm z$ half-spaces described by Eq. \eqref{angdist_FF} that causes the opposite circulation of polarization ellipse orbits for pairs of plane waves radiated into the $\pm z$ half-spaces with equal in-plane Bloch wave vector projections $\mathbf{K}$ as depicted in Fig. \ref{F2}c. As a result, despite the fact that parity-broken 2D monolayer arrays radiate elliptically polarized light, i.e., chiral photons, into specific directions (sometimes referred to as extrinsic chirality), the total radiated optical chirality integrated over all observation angles is strictly zero, as required for radiation produced by an achiral source \cite{moser2025conservation}, i.e., the SLRs considered in this work. Net nonzero radiated optical chirality remains possible for other monolayer arrays provided some of the sites are vertically displaced such that $\hat{\sigma}_h \notin G_\mathbf{0}/T$ \cite{cerdan2023chiral}.  

In addition to its role in determining the polarization state of photons radiated into the far-field, the $\hat{\mathcal{C}}_3$ rotational symmetry of these systems, combined with the fixed phase relationship between LSP dipole moments and their radiated fields, requires the relative phase of photons emitted into directions $\hat{\mathbf{k}}_j$, $j\in\{0,1,2\}$, to be directly inherited from the PAM eigenvalue equation Eq. \eqref{Cn_gen_eig}. The relative phases $\Delta \phi_{j\,j-1}$ imposed by this constraint between pairs of photons radiated into directions $\hat{\mathbf{k}}_j$ and $\hat{\mathbf{k}}_{j-1}$ are tabulated in Fig. \ref{F2}e for each of the $\text{K}$ valley SLRs supported by the parity-broken kagome arrays. \textcolor{black}{Despite present experimental capabilities in the measurement of emission angle, polarization, and phase information connected to optical emission from nanophotonic systems \cite{guo2019lasing, lasingKoenderink2024spontaneous, lasingKoenderink2025metasurface, matsukata2021chiral,zheng2026circularly, machfuudzoh2026cathodol}, valley SLR PAM has not been directly characterized. Yet, we find that it is clearly encoded in the phase profile of the associated optical radiation and should be readily observable, e.g., using Gerchberg-Saxton or other wavefront phase retrieval techniques \cite{gerchberg1972practical}.}


\begin{figure}
    \centering
\includegraphics[width=1.0\linewidth]{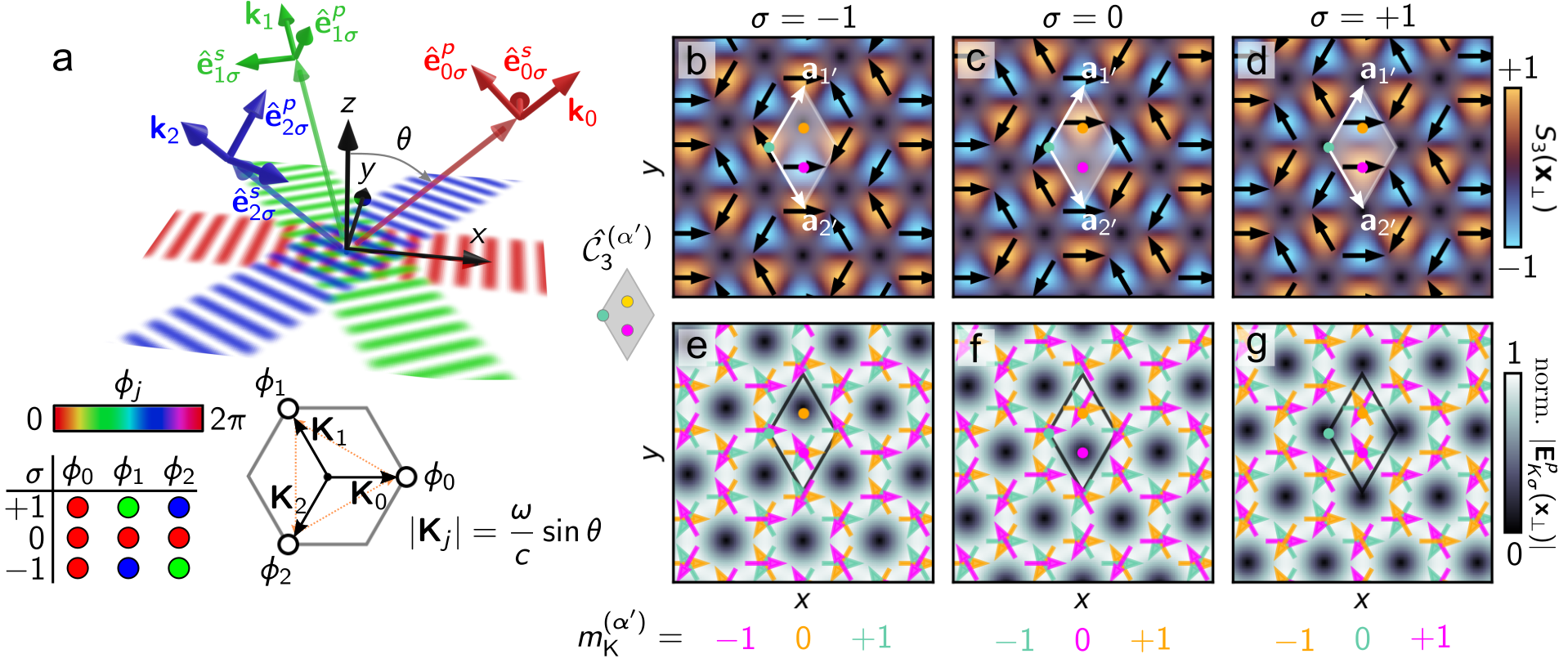}
    \caption{{\bf Optical pinwheel fields with well-defined PAM.} a) Scheme of the optical pinwheel field $\mathbf{E}^\eta_{\text{K} \sigma}$ comprised of three phased plane waves depicted in direct space (upper) and reciprocal space (lower). The value of $\sigma \in \{0,\pm1\}$ dictates the phase $\phi_j$ applied to each plane wave, indicated by color. Plane wave colors in upper panel of (a) correspond to $\sigma=+1$. b)--d) show the spatial variation of the $S_3(\mathbf{x}_\perp)$ Stokes parameter and the antinode field polarization at $t=0$ (black) for $\mathbf{E}^p_{\text{K} \sigma}$ with $\sigma \in \{0, \pm1\}$. e)--g) Normalized $|\mathbf{E}^p_{\text{K} \sigma}|$ profiles. Colored arrows represent the resultant antinode field vectors following $\hat{\mathcal{C}}^{(\alpha')}_3$ rotation of polarization textures in (b)--(d) with respect to each of the color-coded $\alpha'$ axes.}
    \label{F3}
\end{figure}

Inspired by the encoding of valley SLR PAM into the relative phases of light radiated into far-field $\text{K}$ point directions demonstrated in Fig. \ref{F2}e, and aiming to leverage electromagnetic reciprocity, PAM-resolved excitation of valley SLRs by tailored structured optical driving fields is now considered. In analogy to the three-fold phased plane wave fields recently identified in connection to selective excitation of phonons in atomic monolayers with well-defined PAM \cite{bourgeois2025strategy}, we define the optical pinwheel field as
\begin{equation}
    \begin{aligned}
        \mathbf{E}^{\eta}_{\text{K}\sigma}(\mathbf{x}) &= \frac{E_0}{3}\sum_{j=0}^2 \hat{\mathbf{e}}^{\eta}(\mathbf{k}_j, \omega) \  e^{ i(\mathbf{k}_j \cdot \mathbf{x}+ \sigma \phi_j)} \\
        &= \frac{E_0}{3}\sum_{j=0}^2 \bigg\{ \bigg[\hat{\mathcal{C}}^{(\alpha{'}){j}}_3\hat{\mathbf{e}}^{\eta}(\mathbf{k}_0, \omega)  \bigg] \ e^{ i\sigma \phi_j} \bigg\} e^{ i\mathbf{k}_j \cdot \mathbf{x}}  \\
        &= \frac{E_0}{3}\sum_{j=0}^2 \hat{\mathbf{e}}^{\eta}_{j \sigma}(K, \omega)   e^{ i\mathbf{k}_j \cdot \mathbf{x}}  \\
    \end{aligned}
    \label{E_pinwheel}
\end{equation}
where $\phi_j = (2\pi/3)j$, $\mathbf{k}_j = (K \cos{\phi_j}, K \sin{\phi_j}, k_z)$ with $K=|{\bf K}|$, and $\sigma \in \{0, \pm 1\}$. Plane wave polarization vectors are $\hat{\mathbf{e}}^{\eta}_{j \sigma}(K, \omega) = \big[\hat{\mathcal{C}}^{(\alpha{'}){j}}_3\hat{\mathbf{e}}^{\eta}(\mathbf{k}_0, \omega) \big] e^{ i\sigma \phi_j}$ with $\eta \in \{ s,p \}$. Taking $\hat{\mathbf{e}}^{p}(\mathbf{k}, \omega) \equiv \hat{\boldsymbol{\theta}}(\hat{\mathbf{k}})$ and $\hat{\mathbf{e}}^{s}(\mathbf{k}, \omega)  \equiv \hat{\boldsymbol{\varphi}}(\hat{\mathbf{k}})$, the unit polarization vectors become 
\begin{equation}
    \begin{aligned}
        \hat{\mathbf{e}}^{p}_{j \sigma}(K, \omega) &= 
     \begin{pmatrix}
            \cos (2\pi j /3)\cos \theta \\
            \sin(2\pi j /3)\cos \theta  \\
            -\sin \theta
        \end{pmatrix}e^{ i\sigma \phi_j},
    \end{aligned}
    \label{e_hat_jsigmap}
\end{equation}
and
\begin{equation}
    \begin{aligned}
        \hat{\mathbf{e}}^{s}_{j \sigma}(K, \omega) &= 
     \begin{pmatrix}
            - \sin (2\pi j /3) \\
            \cos(2\pi j /3)  \\
            0
        \end{pmatrix}e^{ i\sigma \phi_j}.
    \end{aligned}
    \label{e_hat_jsigmas}
\end{equation}
Fig. \ref{F3}a depicts the optical pinwheel field $\mathbf{E}^p_{\text{K} \sigma}$ in direct space (upper) and reciprocal space (lower). By construction, reciprocal space contributions to pinwheel fields reside on the vertices of an equilateral triangle with $|\mathbf{K}_j| = (\omega/c)\sin \theta$, and the three constituent plane waves interfere in real space to produce a periodic in-plane field profile $|\mathbf{E}^{\eta \perp}_{\text{K}\sigma}(\mathbf{x}, \omega)|$ with hexagonal symmetry defined by optical lattice constant $a' = 4 \pi/3|\mathbf{K}_j|$.

The spatial variation of the in-plane polarization state at fixed $z$ can be characterized using the normalized $S_3$ Stokes parameter $S_3(\mathbf{x}_\perp, z, \omega) = -({2}/{|\mathbf{E}_\perp|^2})\text{Im}\big\{ E_x E^*_y \big\},$ which quantifies local field handedness defined within a 2D plane. Figs. \ref{F3}b--d show $S_3(\mathbf{x}_\perp, z=0, \omega)$ for $\mathbf{E}^p_{\text{K} \sigma}$ and $\sigma\in \{\pm1 , 0\}$ at arbitrary $\omega$. The superimposed black arrows indicate the polarizations of the antinodal field at (arbitrary) time $t=0$, while white diamonds representing the polarization texture unit cells are included with the locations of the three nonequivalent $\hat{\mathcal{C}}^{(\alpha')}_3$ axes marked by color coded circles. It is evident that there are two $\mathbf{E}^p_{\text{K} \sigma}$ anti-nodes per unit cell with equal and opposite circular polarization $S_3 = \pm 1$. These periodic polarization textures are closely related to the topological electromagnetic field textures recently identified in the interferometric patterns created by phased surface plasmon polaritons \cite{tsesses2018optical, du2019deep, dai2020plasmonic, davis2020ultrafast, ghosh2023spin, PhysRevResearch.6.013163}. \textcolor{black}{Similar free-space optical crystals, generalizations of single vortex core beams carrying orbital AM \cite{chong2015polarization, lu2018kaleidoscope, yu2011light}, have already been demonstrated \cite{lin2024optical}.}

Comparison of these panels indicates that the choice of $\sigma \in \{ \pm1 , 0\}$ vertically shifts the periodic field polarization array to cycle the positions of the $\mathbf{E}^p_{\text{K} \sigma}(\mathbf{x}_\perp)$ nodes/anti-nodes around the three $\hat{\mathcal{C}}^{(\alpha')}_3$ rotation axes. Meanwhile, Figs. \ref{F3}e-g show colored arrows representing the antinode field vectors produced following $\hat{\mathcal{C}}^{(\alpha')}_3$ rotation of the black arrow polarization textures in (b)--(d) with respect to each of the color-coded $\alpha'$ axes. In exact analogy to the integer PAM of the valley SLRs supported by parity-broken arrays, the optical pinwheel fields defined by Eq. \eqref{E_pinwheel} exhibit 2D periodic polarization textures with well-defined PAM. The integer PAM $m_\text{K}^{(\alpha')}\in \{\pm1, 0\}$ of $\mathbf{E}^p_{\text{K}\sigma}$ defined with respect to each of the three $\hat{\mathcal{C}}^{(\alpha')}_3$ rotation axes is included below Fig. \ref{F3}e-g for each value of $\sigma \in \{\pm1,0\}$. Therefore, from the perspective of a 2D hexagonal nanoparticle array with $\mathbf{a}_j = \mathbf{a}_j'$, the ability to cycle the pinwheel field $\sigma$ is equivalent to the ability to apply driving fields with well-defined PAM $m^{(\alpha')}_\text{K}=\{\pm1, 0\}.$ 

\begin{figure}
    \centering
    \includegraphics[scale=1.0]{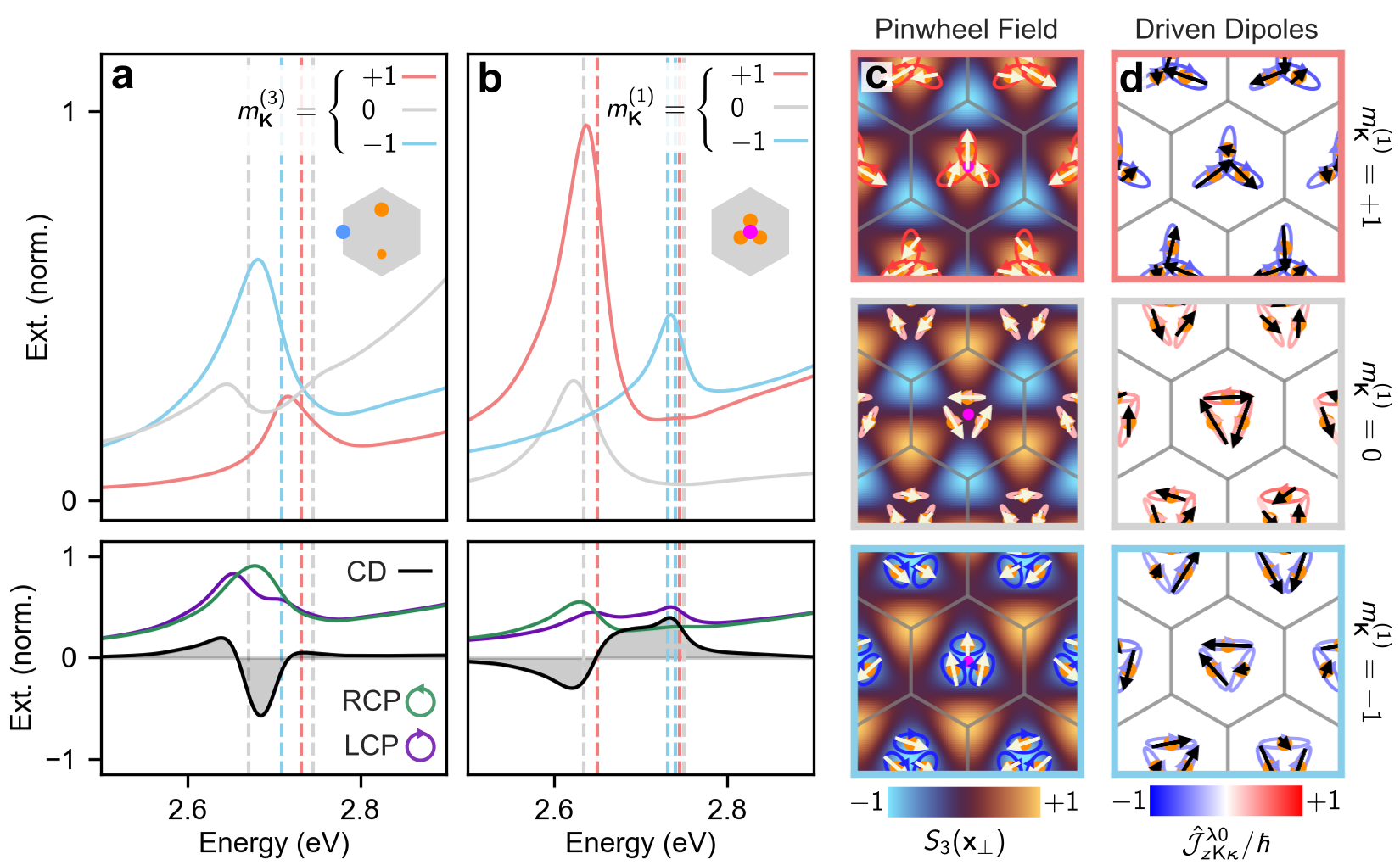}
    \caption{{\bf PAM-resolved SLR measurements with optical pinwheel fields.}
    a) and b) show calculated extinction spectra for the parity-broken honeycomb and kagome arrays considered in Fig. \ref{F1}, respectively, driven by $s$-polarized optical pinwheel fields $\mathbf{E}^s_{\text{K} \sigma}$. Spectra and vertical dashed lines indicating the \textcolor{black}{$\hbar \bar{\omega}_\lambda$} energies are color-coded according to the PAM values in Table \ref{PAM_table}. Green and purple extinction spectra corresponding to right- and left-circularly polarized plane wave excitation are included for comparison, along with the circular dichroism spectrum in black. c) Spatial variation of the $S_3(\mathbf{x}_\perp)$ Stokes parameter and in-plane field polarization vectors at the locations of the parity-broken kagome sites (white arrows) associated with driving optical pinwheel fields $\mathbf{E}^s_{\text{K} \sigma}$ for $\sigma \in {0, \pm1}$. d) Direct-space images of induced LSP dipole polarizations,  colored by $ \hat{\mathcal{J}}^{\lambda 0}_{z\text{K}\kappa} $, under $\mathbf{E}^s_{\text{K} \sigma}$ pinwheel field driving. Box frame colors in (c) and (d) correspond to the driving pinwheel field PAM values $m^{(1)}_{\text{K}}$.}
    \label{F4}
\end{figure}

Extinction spectra of the parity-broken honeycomb and kagome arrays from Fig. \ref{F1}d,e under $s$-polarized pinwheel field $\mathbf{E}^s_{\text{K}\sigma}$ driving with $\theta=\sin^{-1}[K/(\omega/c)]\approx23^\circ$ to match the in-plane Bloch momentum $K$ of the valley SLRs are shown in Fig. \ref{F4}a,b, demonstrating selective excitation of SLRs satisfying $ m^{(1)}_{\lambda \text{K} } = m^{(1')}_\text{K}$. The use of $s$-polarization restricts the coupling to in-plane polarized SLRs, whereas $p$-polarized waves and $\mathbf{E}^p_{\text{K}\sigma}$ generally access both in- and out-of-plane polarized SLRs (SI Fig. S9). PAM selectivity is evident from the progression of resonance peaks in Fig. \ref{F4}a,b, which coincide with the valley SLR energies \textcolor{black}{$\hbar \bar{\omega}_{\lambda}$} (vertical dashed lines) possessing the same PAM as the driving field. Coupling strengths are dictated by the overlap between SLR polarization textures (Fig. \ref{F1}j,k) or those of their radiated fields (Fig. \ref{F2}d) with the driving pinwheel fields (Fig. \ref{F4}c).

Fig. \ref{F4}d shows the LSP dipole polarizations comprising the driven SLRs in the parity-broken kagome lattice at $\hbar\omega$ corresponding to the extinction maxima in Fig. \ref{F4}b. Analogous plots of the driven LSP dipole polarizations in the parity broken honeycomb lattice are included in the Supporting Information (SI Fig. S6). For pinwheel fields with $m^{(1')}_\text{K}=+1$ (light red) and $m^{(1')}_\text{K}=0$ (gray), the driven dipole profiles in Fig. \ref{F4}b closely resemble those of the SLRs with the same PAM value (Fig. \ref{F1}k). However, the $m^{(1')}_\text{K}=-1$ pinwheel drive (light blue) excites a superposition of two nearly-degenerate SLRs carrying the same PAM value, leading to a driven dipole profile that differs from that of either of the underlying SLRs (SI Figs. S2 and S3). It is also worth noting that while the handedness of the driven dipole and pinwheel field polarization ellipse orbits coincide at each site $\mathbf{x}_{\mathbf{n}\kappa}$ when $m^{(1')}_\text{K}=0$ and $m^{(1')}_\text{K}=-1$, they are opposite when $m^{(1')}_\text{K}=+1$. Nevertheless, excitation occurs in all cases only when the PAM of the driving pinwheel field matches the SLR (total) PAM. \textcolor{black}{Examination of the influence of perturbations in relative amplitude and convergence angle upon these selection rules can be found in the Supporting Information (Fig. S10).}

In contrast, the extinction spectra obtained under right- (RCP, green) and left- (LCP, purple) circularly polarized plane-wave excitation, as well as their difference -- the circular dichroism spectra (CD, black), presented in the lower portions of Fig. \ref{F4}a,b, exhibit broadband responses with resonance features spanning multiple SLR energies $\textcolor{black}{\hbar\bar{\omega}_{\lambda}}$ and PAM values $m^{(\alpha)}_{\lambda \text{K} }$. This observation is further supported by the driven dipole polarization profiles induced under circularly polarized extinction, shown in SI Figs. S7 and S8. Therefore, despite the ubiquity of circularly polarized measurements such as CD in the characterization of chiral \cite{fan2012chiral, chiralemissionzhang2022chiral, moser2025conservation} and parity-broken systems \cite{zheng2026circularly}, Fig. \ref{F4}a,b demonstrate their inability to selectively probe SLRs with specific PAM values. Taken together, Fig. \ref{F4} highlights the unique utility of the proposed pinwheel fields in the measurement of PAM-dependent lattice polariton excitations.

The efficient coupling of lattice resonances to light enables access to material valley and AM degrees of freedom through structured optical excitation and polarimetry measurements. However, the relationship between the valley polarization of the nanoparticle LSPs comprising the lattice resonances and the reflection of PAM in the associated electromagnetic fields has remained unclear, limiting selective access and full characterization of valley SLRs. \textcolor{black}{Here, we show that the quantized PAM of K/K'-point SLRs in parity-broken honeycomb and kagome plasmonic arrays is encoded in the polarization, nonlocal phase profile, and angular distribution of their radiative emission, thereby providing direct far-field access to valley polarization degrees of freedom.} We further examine the effects of parity symmetry breaking on SLR PAM and its relation to the AM of the constituent LSPs on each nanoparticle in the array. Building on the connection between the conservative transfer of PAM between SLR and radiation field, we leverage electromagnetic reciprocity to design a new class of structured free-space optical fields, called pinwheel fields, with engineered 2D periodic translation and rotation symmetries, polarization textures, and matching integer PAM content to selectively drive valley-polarized SLRs. \textcolor{black}{The connections established between common symmetries reflected within lattice excitation and radiation field polarization textures provide new opportunities for harnessing valley polarization to encode, transport, and process information in structured optical cavities and quantum devices beyond traditional chiral light signals \cite{liu2020generation, liu2021valley, kang2018pseudo, dong2017valley, liu2020z2, shao2023topological, chen2020chiral, tao2023multi}.}

\textcolor{black}{\section{Methods}
All hexagonal lattices considered in this work are composed of oblate-spheroid-like silver nanoparticles surrounded by vacuum. The bulk silver dielectric response is described by a Drude free electron gas model with plasma frequency $\hbar\omega_p =8.9$ eV, damping rate $\hbar\gamma =0.0387$ eV, and high-frequency limit $\varepsilon_{\infty}=5$, as reported in Ref. \cite{yang2015optical}. Intrinsic SLR energy-momentum dispersions $\hbar \omega_\lambda(\mathbf{K})$ and polarizations $\mathbf{p}^\lambda_{\mathbf{K}\kappa}$, as well as driven responses under individual plane wave and structured pinwheel driving fields, are evaluated using the coupled dipole model \cite{cherqui2019plasmonic, kravets2018plasmonic, Markel1993, Zou2004_1, Manjavacas2018}, described in more detail in the Supporting Information. The optical response of each nanoparticle is characterized by a diagonal anisotropic dipole polarizability tensor with elements $\alpha_{\beta \beta}(\omega)=i(3/2k^3)a_1(\omega)$, where $a_1(\omega)$ is the $\ell=1$ electric Mie coefficient for a sphere of radius $R$ \cite{bohren2008absorption}. The in-plane $\beta=x,y$ and out-of-plane $\beta=z$ polarizability components correspond to $R=100$ nm and $R=60$ nm, respectively.}

\begin{acknowledgement}
This research was supported by the National Science Foundation under Award No. CHE-2516408.
\end{acknowledgement}

\newpage
\bibliography{refs}
\end{document}